\documentclass[preprint]{aastex}
\usepackage{color}
\usepackage{graphicx}
\usepackage{amssymb}
\usepackage{amsmath}
\usepackage{appendix}

\shorttitle{VLT/X-shooter spectroscopy of MAXI~J1659--152}
\shortauthors{Kaur et al.}

\begin{document}

\title{VLT/X-shooter spectroscopy of the candidate black-hole X-ray
binary MAXI~J1659--152 in outburst\footnote{Based on  
ESO-VLT/X-shooter  observations obtained using the X-shooter 
guaranteed time GRB program (086.A-0073) }}

\author{Ramanpreet Kaur\altaffilmark{1,2}, 
Lex Kaper\altaffilmark{1},
Lucas E. Ellerbroek\altaffilmark{1}, 
David M. Russell\altaffilmark{1}, 
Diego Altamirano\altaffilmark{1}, 
Rudy Wijnands\altaffilmark{1}, 
Yi-Jung Yang\altaffilmark{1},
Paolo D'Avanzo\altaffilmark{3}, 
Antonio de Ugarte Postigo\altaffilmark{4,8}, 
Hector Flores\altaffilmark{5}, 
Johan P.U. Fynbo\altaffilmark{3}, 
Paolo Goldoni\altaffilmark{6,7}, 
Christina C. Th$\ddot{\rm o}$ne\altaffilmark{8,9}, 
Alexander van der Horst\altaffilmark{1,10},
Michiel van der Klis\altaffilmark{1},
Chryssa Kouveliotou\altaffilmark{11},
Klaas Wiersema\altaffilmark{12}
and Erik Kuulkers\altaffilmark{13}}
  
\altaffiltext{1}{Astronomical Institute `Anton Pannekoek', University
  of Amsterdam, Science Park 904, 1098 XH Amsterdam, the Netherlands}
\altaffiltext{2}{Physics Department, University of
  Wisconsin-Milwaukee, Milwaukee, WI 53211, USA}
\altaffiltext{3}{INAF, Osservatorio Astronomico di Brera, via
  E. Bianchi 46, 23807 Merate (LC), Italy} 
\altaffiltext{4}{Dark Cosmology Centre, Niels Bohr Institute, Juliane
  Maries Vej 30, Copenhagen, 2100, Denmark} 
\altaffiltext{5}{GEPI, Paris Observatory, CNRS, University of
  Paris-Diderot; 5 Place Jules Janssen, 92195 Meudon, France}
\altaffiltext{6}{Laboratoire Astroparticule et Cosmologie, 10 rue
  A. Domon et L. Duquet, 75205 Paris Cedex 13, France}
\altaffiltext{7}{Service dÕAstrophysique, DSM/IRFU/SAp, CEA-Saclay, 91191 Gif-sur-Yvette, France}
\altaffiltext{8}{IAA - CSIC, Glorieta de la Astronom$\acute{\rm i}$a
  s/n, 18008 Granada, Spain} 
\altaffiltext{9}{Niels Bohr International Academy, Niels Bohr
  Institute, Blegdamsvej 17, 2100 Copenhagen, Denmark}
\altaffiltext{10}{USRA, NSSTC, Huntsville, AL 35806, USA}
\altaffiltext{11}{Space Science Office, VP62, NASA/Marshall Space Flight
Center, Huntsville, AL 35812, USA}
\altaffiltext{12}{Department of Physics and Astronomy, University of
  Leicester, University Road, Leicester LE1 7RH, UK}
\altaffiltext{13}{European Space Agency, European Space Astronomy Centre, P.O. Box 78, 28691, Villanueva de la Ca$\tilde{\rm n}$ada, Madrid, Spain}
\email{r.kaur@uva.nl}

\begin{abstract}
 We present the optical to near-infrared spectrum of MAXI~J1659--152, during the onset of its 2010 X-ray outburst.
 The spectrum was obtained with X-shooter on the ESO {\it Very Large Telescope} (VLT) early in the 
  outburst simultaneous with high quality observations at both shorter 
 and longer wavelengths. 
At the time of the observations, the source was in the low-hard state.  
  The X-shooter spectrum includes many broad ($\sim
  2000$~km~s$^{-1}$), double-peaked emission profiles of H, He~{\sc i}, He~{\sc ii},  
  characteristic signatures of a low-mass X-ray binary during outburst. 
    We detect no spectral signatures
  of the low-mass companion star. 
  The strength of the diffuse
  interstellar bands results in a lower limit to the total
  interstellar extinction of $A_{V}$ $\simeq$ 0.4~mag. Using the neutral 
  hydrogen column density obtained from the X-ray spectrum 
  we estimate $A_{V} \simeq 1$~mag. The radial-velocity structure of the
  interstellar Na~{\sc i}~D and Ca~{\sc ii} H\&K lines results in a
  lower limit to the distance of $\sim 4$ $\pm$ 1~kpc, consistent with
  previous estimates. With this distance and $A_{V}$, the  dereddened  spectral
  energy distribution represents a flat disk spectrum. 
  The two 10~minute X-shooter spectra show
  significant variability in the red wing of the emission-line
  profiles, indicating a global change in the density structure of the
  disk, though on a timescale much shorter than the typical viscous
  timescale of the disk.
\end{abstract}


\section{Introduction}
 
Transient low-mass X-ray binaries (LMXBs) form a sub-class of compact
binary systems in which a neutron star or a black hole accretes matter
from a low-mass companion through  Roche-lobe overflow and occasionally 
undergo X-ray outbursts, during which their 
X-ray flux increases by several orders of magnitude in
comparison to the quiescent state (\citealt{Tanaka1995xrbi}; \citealt{vanParadijs1995}). 
These outbursts have been observed on the timescales from a few
weeks to several years. In outburst the X-ray emission of these systems is
directly related to the accretion process and originates from the inner accretion disk
while the optical emission  is dominated
from the reprocessed emission from the accretion
disk with some contribution from the heated surface of the companion
star. Many transient LMXBs have been studied in detail in the X-rays, but for
only a few systems simultaneous high quality optical spectroscopy has
been obtained during their outburst (e.g. GX~339--4: \citealt{Soria1999}, XTE~J1118+480:
\citealt{Dubus2001}).  The latter is important to study the large-scale
physical structure of the accretion disk.

On September 25, 2010, at UT 08h05, the {\it Burst Alert Telescope}
(BAT; \citealt{Barthelmy2005}) onboard the {\it Swift} satellite
was triggered by a transient source, first thought
to be a gamma-ray burst: GRB 100925A \citep{Mangano2010}.  On the
same day, the {\it Monitor of All-sky X-ray Image} (MAXI) satellite
\citep{Matsuoka2009} also reported the discovery of a new transient,
named MAXI~J1659--152 (hereafter MAXI1659), at a position consistent
with the GRB \citep{Negoro2010}. Subsequently, a number of
space-based and ground-based observations were carried out to
investigate the nature of this source (e.g. \citealt{vanderHorst2010},  \citealt{Vovk2010}).  Immediately after the BAT
trigger, the {\it UltraViolet/Optical Telescope} (UVOT) onboard the \textit{Swift} satellite  \citep{Roming2005} 
detected the optical counterpart of MAXI1659 with a magnitude of 16.8 in the white filter
\citep{Marshall2010}.
  
On September 25 at UT 23h39, the X-shooter spectrograph on the ESO
{\it Very Large Telescope} (VLT) was pointed at the optical
counterpart and registered an optical to near-infrared spectrum of
MAXI1659 revealing the Galactic origin of the source \citep{deUgartePostigo2010}. 
The X-shooter spectrum showed various broad double peaked
emission line profiles, as typically observed in X-ray binaries when they actively accrete.

Subsequent studies of the X-ray spectral and timing behaviour of the
source (\citealt{Kalamkar2011}, \citealt{Kennea2011}, \citealt{MunozDarias2011})
demonstrated that the system is a candidate black-hole X-ray
binary. MAXI1659 also shows regular, though not really periodic dips
in the X-ray intensity with a period of approximately 2.42h (\citealt{Kennea2011}, 
\citealt{Kuulkers2011}), pointing at a high orbital
inclination. This is also interpreted as an orbital period and makes  MAXI1659 
the shortest orbital period black-hole X-ray
binary candidate with an estimated distance of $\sim$ 7 kpc and possibly an M5
dwarf companion (\citealt{Kuulkers2010}, \citealt{Belloni2010b}, \citealt{Kuulkers2011}). 
Later on, the distance was suggested to be  within 1.6 - 4.2
kpc \citep{MillerJones2011}.  Like some other transient LMXBs
(e.g. Swift J1753--0127, XTE J1118+480, GRO J0422+32, cf.\ \citealt{Zurita2008} )
MAXI1659 is at a high Galactic latitude ({\it b} = +16.49$^{\circ}$).

In this paper, we report on the optical to near-infrared
VLT/X-shooter spectra of MAXI1659 taken during the beginning of its
2010 X-ray outburst (see \citealt{deUgartePostigo2010} for a
preliminary report).

\section{Observations and data reduction}

We observed the optical counterpart of MAXI1659 on September 25,
2010, starting at UT 23h39 (MJD 55464.99) after triggering the
X-shooter guaranteed time GRB program (086.A-0073, PI Fynbo) on the
VLT. X-shooter is a wide-band, medium-resolution spectrograph,
consisting of 3 arms. Each arm is an independent cross-dispersed
echelle spectrograph with its own slit unit. The incoming light is
split using dichroics, resulting in three spectral ranges: UVB
(3000--5900~\AA), VIS (5500--10200~\AA) and NIR (10000--24800~\AA). A detailed
description of the instrument is provided by
\citet{D'Odorico2006} and \citet{Vernet2011}.

We obtained two  spectra of MAXI1659 of 600s, starting at UT
23h39 and UT 23h50, at two different nodding positions on the slit. A
slit width of 1.0\arcsec\ was used for the UVB arm while a 0.9\arcsec\ slit
was used for the VIS and NIR arm; the corresponding resolving power is
R~$\sim$~5100, 7500, and 5700, respectively.  A telluric standard star
(HIP083448) was observed immediately after the MAXI1659
observations. The spectrophotometric standard (Feige~110) was observed
the night before, with the same slit configuration as used for
MAXI1659. The seeing was 1.8\arcsec\ in the {\it V}~ band.

We used the ESO X-shooter pipeline version 1.2.0 (\citealt{Goldoni2006},
\citealt{Modigliani2010}) to obtain the wavelength calibrated
and rectified  two-dimensional spectra.  The observations were reduced 
both in nodding and in staring mode, where in the latter
case the two exposures were reduced separately.  The one-dimensional 
spectrum was extracted and normalized using the standard {\it IRAF} routines {\tt apall} 
and {\tt continuum}. The resultant spectrum was corrected for 
the Earth's motion with respect to the local
standard of rest (LSR) and the NIR spectrum was corrected for telluric absorption lines. The complete X-shooter spectrum was flux
calibrated order-by-order relative to the flux table of spectrophotometric standard Feige~110.
The obtained signal to noise ratio for the combined spectra is 46, 37
and 8 in the UVB, VIS and NIR arm, respectively.

We also analysed the archival optical and ultraviolet observations of
MAXI1659 obtained with UVOT onboard the \textit{Swift} satellite starting on September 26,
2010 at UT 0h07, 8 minutes after the X-shooter observations (see also \citealt{Kennea2011}). 
The image data of
each of the six filters used were summed using {\tt uvotimsum} and photometry of the
source in individual sequences was derived with {\tt
uvotsource}. MAXI1659 is clearly detected at a significance of $> 50
\sigma$ in all six filters (1900--5500~\AA). Three trial extraction region radii of  3\arcsec,
4\arcsec, and 5\arcsec were used, and the magnitude  in each filter
were derived from the range of output values.

\section{Results}

\subsection{Emission lines}

The X-shooter spectrum of MAXI1659 includes a number of  emission lines
of H, He~{\sc i} and He~{\sc ii}. 
Fig.~1 displays the normalized X-shooter spectrum of MAXIJ1659 in the range 3000-10000~\AA~
where all identified spectral lines are labelled at their respective wavelengths. 
Among the hydrogen lines, we detect the Balmer series from
H$\alpha$ up to H$\eta$ and the Paschen series from Pa$\beta$ up to
Pa--11. Fig.~2 displays the three strongest
line profiles on a velocity scale showing a double-peaked
structure.
Table 1 lists all the spectral lines detected  along with the
line-profile parameters, measured using the {\it IRAF} routine {\tt splot}.

All emission lines are very broad: the full width at half maximum
(FWHM) is in the range of 1500--2000~km~s$^{-1}$. The peak separation
is measured as the difference in the central position of two gaussians
fitted to the wing of each individual component; for the weaker
emission-line profiles, it is difficult to determine whether they are
also double-peaked. The peak separation varies from 918 to
1175~km~s$^{-1}$, with a clear increasing trend among the H Balmer
lines, except for H$\alpha$ (see Fig.~3). A similar trend in the
peak-to-peak separation has been observed in GX 339--4 but among 
H, He~{\sc i} and He~{\sc ii} lines  \citep{Wu2001}. Such a trend
indicates that higher excitation lines show a larger peak separation,
i.e. they are formed closer to the black hole where the disk is hotter
and more rapidly rotating. However, in our spectra, the He 
lines do not seem to follow this trend (Fig.~3). A possible
explanation may be that the He  lines are predominantly produced near the
hot spot (see below). The  disk velocity of the system is 156 $\pm$ 42~km~s$^{-1}$, measured 
as the difference of the velocity of the center of the two peaks of a line with respect to its rest velocity.
Only the strongest emission lines are used to measure the disk velocity.

In a number of LMXBs, a Bowen blend is detected at $\lambda$
4650--4660~\AA, usually attributed to the high excitation lines of the
N~{\sc iii} $\lambda$4634--4641~\AA /C~{\sc iii}
$\lambda$4647--4650~\AA\ formed due to the Bowen fluorescence
mechanism (BFM, \citealt{McClintock1975}, \citealt{Steeghs2002};
\citealt{Casares2006}).  In MAXI1659 we detect a broad but weak emission line feature centered at
4628~\AA\ with a FWHM of 2640~km~s$^{-1}$ which is likely due
to the Bowen blend. However we detect no narrow
peaks/structures corresponding to the individual N~{\sc iii}
components in this blend. The central velocity of this feature is --775 ~km~s$^{-1}$, 
which is very different from the disk velocity of the system (156 km~s$^{-1}$)  
 and may be related to the disk wind.  In another black hole binary XTE J1118+480, the Bowen 
blend was detected as a weak feature during outburst but was observed to vary in phase with He~{\sc ii} 4686 \AA~ \citep{Dubus2001}.~ 
Assuming that we observe the varying peak of the Bowen blend, we would expect to 
detect the red, rather than the blue component as the hydrogen and helium lines show  the variability in the red component (see Section 3.2).

\subsection{Emission-line variability}

During the two 10 minute-observations of MAXI1659, the red
wing of the emission-line profile shows significant variability
(cf. Fig.~2). During the two observations, the FWHM and the peak
separation remain  constant while the equivalent width (EW) of the lines increases (e.g.,
the EW of H$\alpha$ increases from --7.3$\pm$0.4 to --8.3$\pm$0.4~\AA, cf.\ Fig.~2). 
This variability is detected in different lines at the same velocity. As these lines originate in different parts of the spectrum (even in different spectrograph arms) this excludes the possibility that the observed variations are of instrumental nature.

Our X-shooter observations sample the orbital period of the source in which strong dipping is expected as estimated using 
an X-ray dip activity ephemeris (Kuulkers et al. 2012 in prep). The lack of simultaneous X-ray observations and, the fact that  
dips in this source show highly irregular structure which can last from a few minutes to as long as $\sim40$ min (Kuulkers et al. 2011, Kuulkers et al. 
2012 in prep), does not allow us to know which percentage of the X-shooter observations is affected by dips.
However,  the observed variability in MAXI1659 is similar to that observed in XTE~J1118+480
\citep{Dubus2001}; in this system the He~{\sc ii} 4686~\AA\ line
time series displays an S-wave pattern consistent with the photometric
(i.e. orbital) period of 4.1h. According to Dubus et al. the variable
component of the He~{\sc ii} line (as well as that observed in
H$\alpha$, H$\beta$ and the Bowen blend) most likely originates in the
hot spot of the accretion disk where the accretion flow impacts on the
disk. 

\subsection{Interstellar spectrum and distance to MAXI1659}

The spectrum includes a number of interstellar lines: the Na~{\sc i} D
and Ca~{\sc ii} H\&K lines, and a few diffuse interstellar bands
(DIBs); their EW is listed in Table 2. \citet{Cox2005} measured the DIB spectrum in the strongly reddened
sightline towards the  X-ray binary 4U1907+97 ($E(B-V)$ = 3.45)
and compared the EWs of several DIBs to those observed in the
well-studied sightlines towards BD+631964 ($E(B-V)$ = 1.01) and HD183143
($E(B-V$) = 1.28). We calculated the EW ratio of the DIBs detected in
MAXI1659 to those in the three mentioned sightlines, and arrive at a
ratio of 0.06, 0.13 and 0.08, i.e. an $E(B-V)$ estimate of 0.21, 0.13
and 0.10~mag, respectively. As we do not know the value of $R_{V}$ in
the line of sight to MAXI1659, we used the $R_{V}$ value for the respective
sightlines (2.75, 3.1, and 3.3), and by using $A_{V}$ = $E(B-V$)$*R_{V}$
we estimate $A_{V}$ as 0.57, 0.40, and 0.33~mag, respectively. Thus,
based on the DIB spectrum, $A_{V}$ is  $0.4 \pm 0.1$~mag.

Adding the {\it Swift}-UVOT data to extend the spectral coverage of
the X-shooter spectrum towards the ultraviolet, we note a broad 
feature at 2175~\AA\  (``the extinction bump'') which suggests a higher 
value of the optical extinction (see Section 3.4). Given the height of
MAXI1659 above the Galactic plane ($z \sim 1$~kpc, see below), it may
well be that the (still unidentified) DIB carriers are concentrated towards
the Galactic plane, thus providing a lower limit to the total extinction.

The EW of the DIB at 5780~\AA\ can be related to the neutral hydrogen
column density $N_{H}$; using Cox et al. (2005) a rough estimate of
$N_{H}$ = 0.6 $\times$ 10$^{21}$~cm$^{-2}$ is obtained. Given the
above, we also expect this to be a lower limit to the total $N_{H}$
value. {\it Swift} X-ray observations \citep{Kennea2011} indicate
that the interstellar $N_{H}$ value is $2.4 \times 10^{21}$~cm$^{-2}$.

An absorption feature is
present in the red wing of the H$\beta$ emission line profile. This
feature has been observed in other systems (e.g. \citealt{Buxton2003}, 
\citealt{Soria2000}, \citealt{Dubus2001}) and has been
interpreted as a redshifted absorption component in H$\beta$; however,
we think it most likely is due to the strong DIB at 4882~\AA~\citep{Jenniskens1994}.

The Na~{\sc i} D and Ca~{\sc ii} H\&K interstellar lines are centered
at a velocity of $14 \pm 10$~km~s$^{-1}$ with respect to the LSR of the solar environment. A kinematic model of
the differential Galactic rotation \citep{Brand1993} predicts a
steady increase of the radial velocity in the direction of MAXI1659
from 0~km~s$^{-1}$ to 122~km~s$^{-1}$ at 8~kpc. 
The measured radial velocity distribution implies a lower limit to the
distance of MAXI1659 of $4 \pm 1$~kpc. The corresponding height above
the Galactic plane is $z \geq 1$~kpc.

\subsection{Spectral energy distribution} 

Fig.~4 shows the flux-calibrated X-shooter spectrum of the optical
counterpart of MAXI1659 extended to the ultra-voilet region using the {\it Swift}-UVOT 
photometry.  Due to estimated slitlosses, the X-shooter spectrum was scaled up by a factor of 1.75. 
The thus obtained SED is consistent with the UVOT photometry.

The neutral hydrogen column density increased during the outburst due
to absorption intrinsic to the source \citep{Kennea2011}; the
initial, and thus likely the interstellar value was $N_{\rm H} = (2.4
\pm 0.3) \times 10^{21}$ cm$^{-2}$ on MJD~55464 at the start of the
outburst. This corresponds \citep{Guver2009} to a visual extinction
of $A_{V} = 1.1 \pm 0.2$~mag. This value is consistent 
with the value for $A_V$ necessary to remove the extinction bump at 2175~\AA~  in the UVOT photometry.\footnote{
We  note that a slightly higher $A_V$ ($\sim 1.6$ mag) is found by
fitting only the UVOT and optical X-shooter SED with a power-law
reddened with the \citealt{Cardelli1989} extinction law. This different approach will be
presented in a forthcoming companion paper (D'Avanzo et al., in preparation).
}
The X-shooter spectrum and UVOT
photometry are de-reddened adopting this value and using the extinction laws
of \citet{Cardelli1989} and \citet{Mathis1990}. The
dereddened spectral energy (SED) distribution becomes flat with a continuum
rising to the blue as expected for LMXBs in outburst, as shown in red color in Fig.~4.

The whole UV--optical--NIR spectrum can be fitted by a single power
law with slope $\beta$ = -1.37 $\pm$ 0.14 (where $\lambda F_{\lambda}
= \lambda^{\beta}$). The spectrum cannot be fitted by a single
temperature blackbody. For an optically thick, non-irradiated disk, we
expect a power law with an index of $\beta \sim -1.33$ describing a
multi-temperature blackbody extending from the optical/NIR regime to
the far-UV  \citep{Frank2002}. The intrinsic spectrum of
MAXI1659 is remarkably consistent with this canonical spectrum. For
most LMXBs in outburst, $\beta$ ranges from -2.5 to -1.5 as observed from their optical/NIR spectra 
and originates in the irradiated outer disk (e.g. \citealt{Hynes2005}). There is no significant
variation in the spectral slope or colour of the UV--optical SED
during the outburst \citep{Kennea2011}. This suggests that the
non-irradiated disc  dominates the emission throughout the entire outburst
of MAXI1659.  

\section{Discussion}

The optical to near-infrared spectrum of MAXI1659, taken during the
beginning of the 2010 X-ray outburst, is consistent with a
black-hole LMXB  in the low-hard state (see \citealt{Kennea2011} for the
determination of the source state at the time of our X-shooter observations). The spectrum
includes a number of double-peaked emission profiles revealing the
fast rotation of the accretion disk. The observed trend in peak separation of the
hydrogen Balmer line with excitation energy reveals the rotation profile of
the accretion disk, where the inner and hotter regions of the disk
rotate faster than the outer regions \citep{ Smak1981}. 

Although we obtained only two spectra, a significant increase is
observed in the red peak of the emission lines on a ten-minute
timescale. This timescale is much shorter than the timescale
associated with the outburst and the viscous timescale of the
disk, estimated of the order of days to weeks  for typical accretion disk parameters \citep{Frank2002}. 
It is possible that  
the observed change in the double-peaked
emission lines is related to an S-wave pattern as observed in XTE
J1118+480 \citep{Dubus2001} caused by the revolving hot spot where
the accretion stream from the low-mass companion star impacts onto the
disk. This may then also explain why the He lines do not follow the
same trend in varying peak separation as the H Balmer lines. An
alternative explanation for the observed line-profile variability
could be the warping of the disk propagating with time \citep{Maloney1997}. Another possibility 
could  be  the presence of a revolving structure in the accretion 
disk which has also been proposed as a possibly cause of  the X-ray 
dipping behavior (\citealt{DiazTrigo2006}, \citealt{DiazTrigo2009}, \citealt{Boirin2005}). 
  
The SED extending from the far-ultraviolet
down to the near-infrared  is flat when dereddened with $A_{V} =
1.1$~mag, consistent with an optically thick, non-irradiated accretion
disk (Fig 4.). The visual extinction is consistent with that derived from the
2175~\AA\ feature and the observed $N_{H}$ from the X-ray observations. The DIBs
indicate a smaller extinction, suggesting that the (unknown) DIB
carrier(s) are concentrated towards the Galactic plane, apparently
unlike the carrier of the 2175~\AA\ feature. A supporting argument is
that MAXI1659 is located far above the Galactic plane ($\sim 1$~kpc
for a distance of 4~kpc), like the other very few LMXBs   \citep{White1996}.
which are possibly ejected out of the Galactic plane due to large kick 
received during the supernova explosion; alternatively, they are remnants of the population of massive stars formed during 
early stages of the evolution of the Galaxy \citep{Mirabel2001}. 

The measured spectral slope is similar to that observed in the
black-hole LMXB XTE J1118+480: $\beta \sim -1.6$  \citep{Dubus2001}. 
In this source, synchrotron emission from the compact jet
is known to contribute a large fraction of the optical/NIR flux in the
low-hard state, making the spectrum redder than most usually reported
for LMXBs (e.g., 
\citealt{Hynes2003}). It is plausible that the
jet of MAXI1659 detected at radio frequencies (e.g. \citealt{vanderHorst2010}) 
could contribute to the X-shooter spectrum of MAXI1659 as
the source was in the low-hard state during the observation, which is
commonly associated with compact jets seen in the optical/NIR
(e.g. \citealt{Buxton2004}; \citealt{Russell2006}; \citealt{Coriat2009}). It would be interesting 
to monitor the system during an outburst  to study the disk revolution and longer -term 
evolution in response to the outburst. It will remain very difficult to detect the low-mass companion 
star and/or measure the mass of the black hole.

\section*{Acknowledgements} 
RK, RW and JPUF acknowledge support from the European Research Council starting grant. 

\begin{figure*}
\centering
\medskip
\includegraphics[height= 18cm]{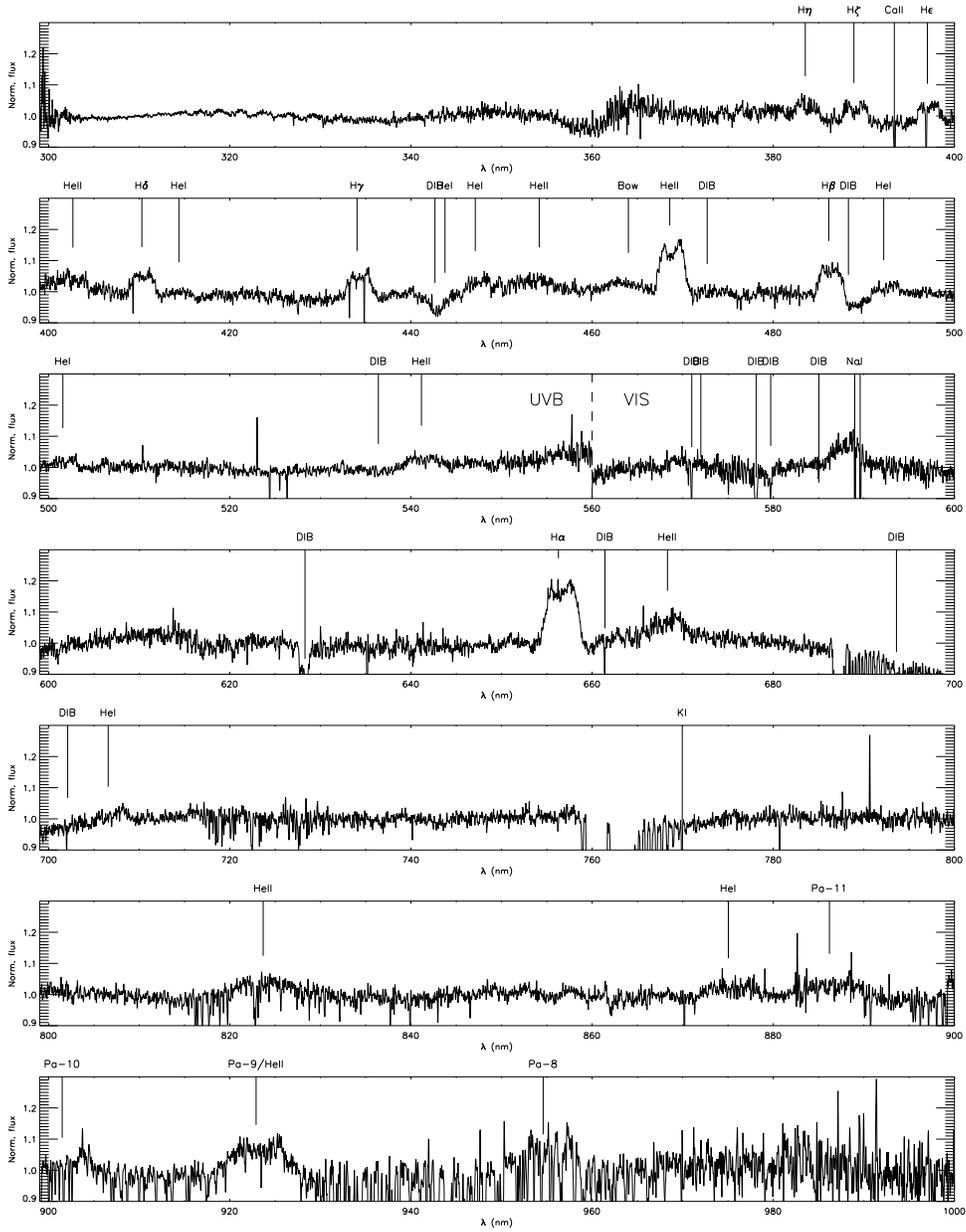}
\caption{The X-shooter normalized spectrum of MAXI J1659-152 in the wavelength range 3000 - 10000~\AA, with all  
identified spectral lines labelled at their respective wavelengths.}
\end{figure*}

\begin{figure*}
\centering
\medskip
\includegraphics[height= 15cm]{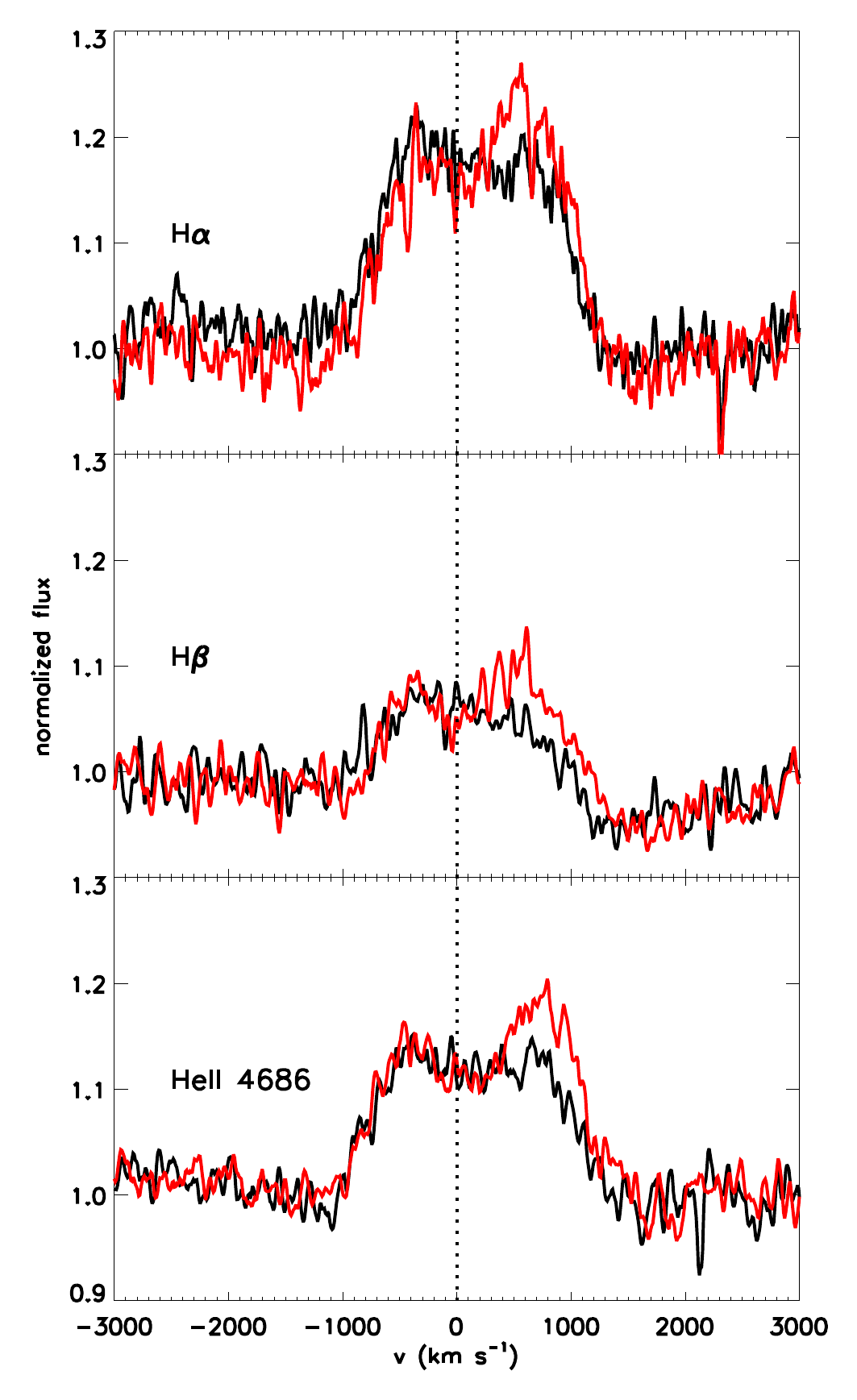}
\caption{The observed double-peaked emission lines in the spectrum of
  MAXIJ1659 on a velocity scale (black : observation 1 -  Sep 25 UT 23:39, red : observation 2 - Sep 25 UT 23:50). Note the variable red wing
  becoming significantly stronger on a timescale of 10 minutes. Zero on the velocity scale corresponds to the local standard of rest.}
\end{figure*}

\begin{figure*}
\centering
\medskip
\includegraphics[height= 15cm]{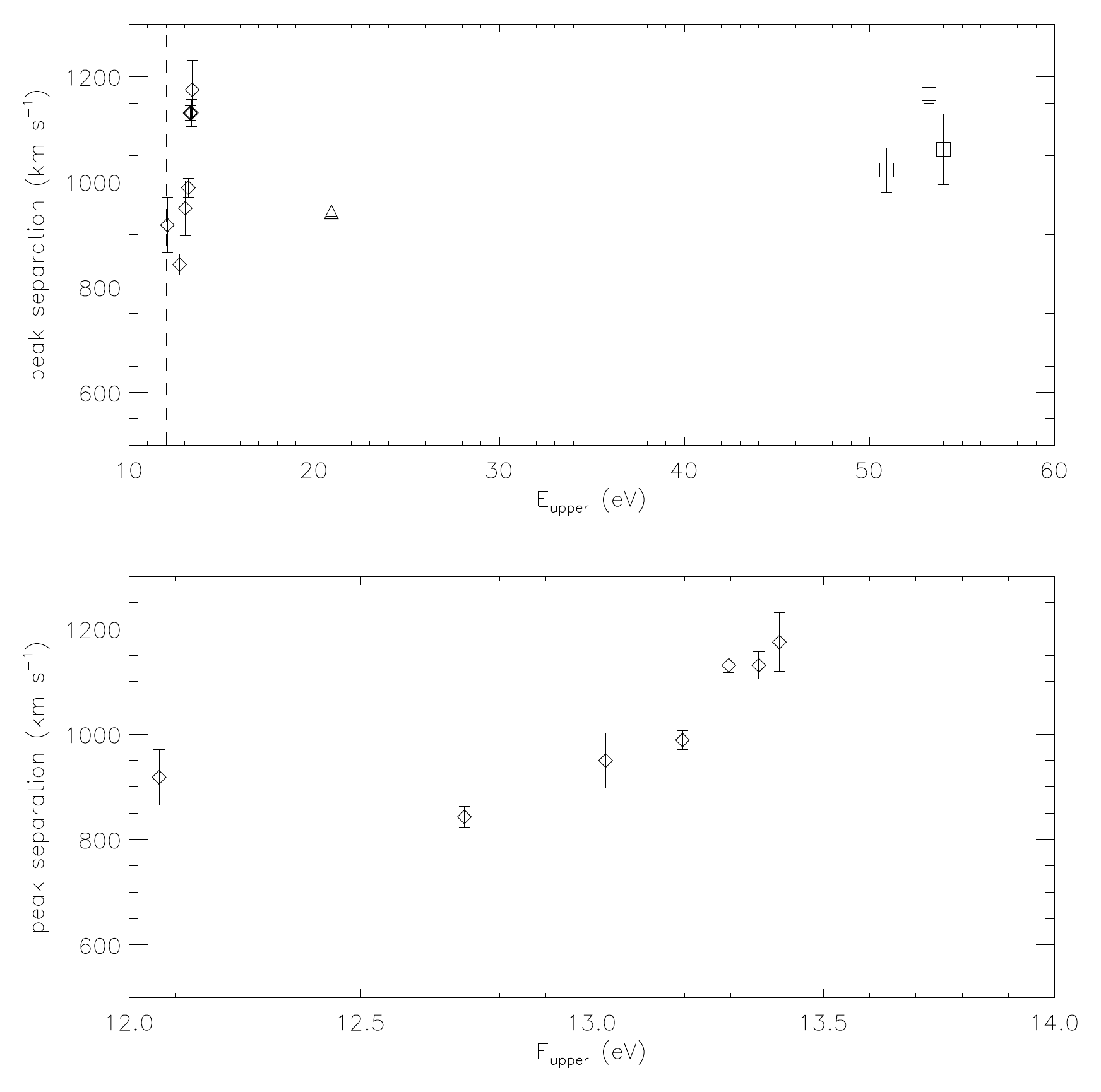}
\caption{ Peak separation versus the energy of the upper level for transitions of H I (diamonds), He I (triangles) and He II (squares). 
The region between dashed lines is enlarged in the lower panel.}
\end{figure*}

\begin{figure*}
\centering
\medskip
\includegraphics[height= 15cm]{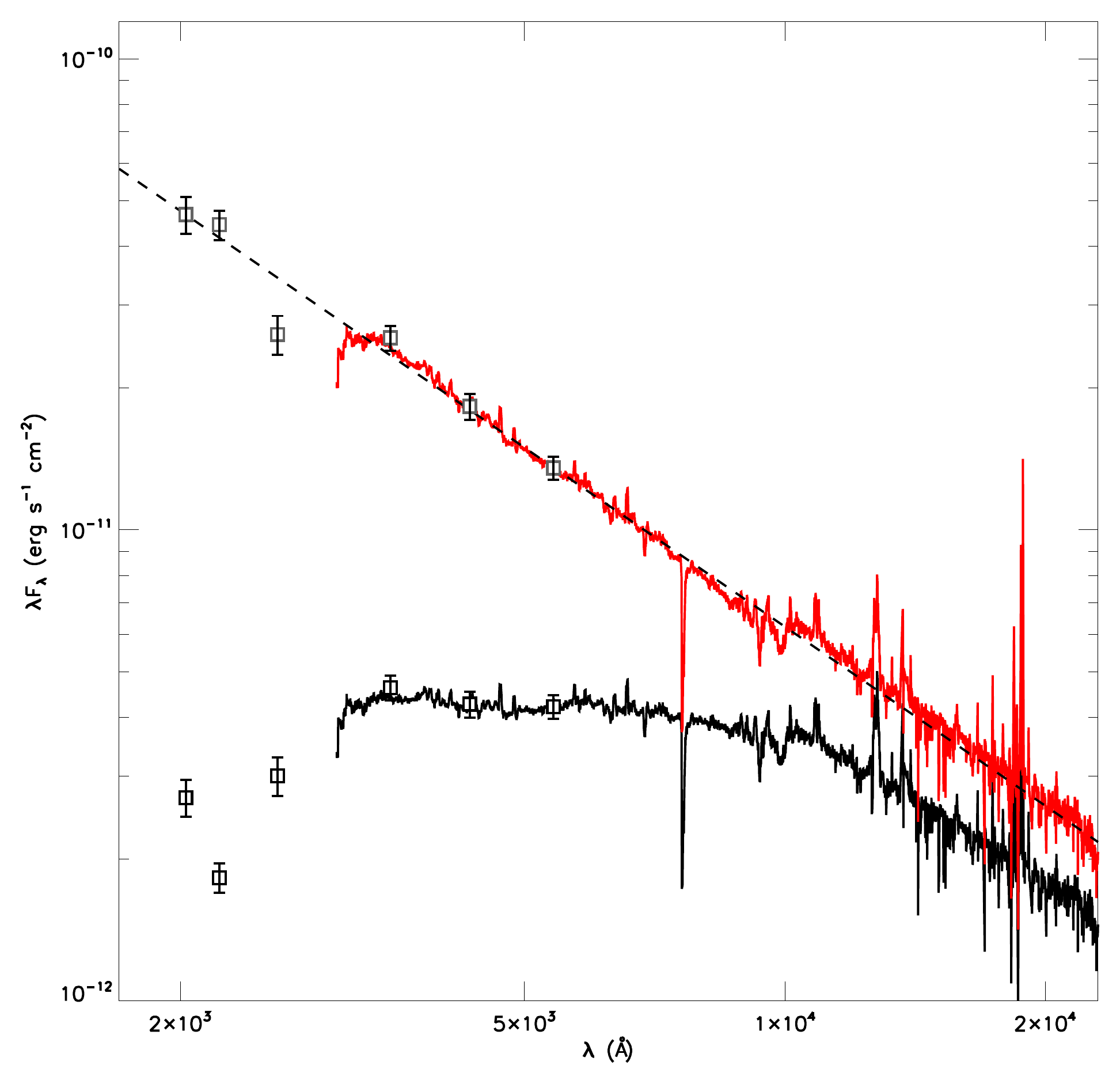}
\caption{MAXI1659 flux-calibrated spectrum (black); dereddened spectrum, Av=1.1 mag (red); {\it Swift}-UVOT photometry (squares). 
The spectrum was rebinned on 5~\AA~ and some features were clipped in order to show the general shape of the SED. The dotted line represents fit to the de-redenned spectra 
with a power-law index of  -1.37.}
\end{figure*}

\newpage

\begin{table}
\centering
\caption{Parameters of spectral lines detected in the X-shooter
  spectrum of MAXI1659. The first column lists the identification and rest
  wavelength of the spectral line,  second column the equivalent
  width,  third column the full width at half maximum and the
  fourth column the peak separation.  A line is marked with `B' in case of a blend.}
\medskip
{%\tiny
\begin{tabular}{lccc}
\hline
\hline
Line	($\lambda_o$)	 	&EW 	& 	FWHM 	&    peak-separation  \\
		~~~~~~~(\AA)	& (\AA)			   &	(km~s$^{-1}$)			&(km~s$^{-1}$) \\			
\hline
H$\eta$ 	3835.38		&	-0.98 $\pm$  0.09		& 	1895 $\pm$ 87  	&	1175 $\pm$ 56	\\
H$\zeta$  	3889.05		&	-1.18 $\pm$  0.03		&	2035	 $\pm$ 69		&	1131	 $\pm$ 26	\\
H$\epsilon$ 3970.07	&	-0.70 $\pm$ 0.09		&	1618 $\pm$ 129	&  	1131 $\pm$ 14	 \\		
H$\delta$ 	4101.73		&	-1.57 $\pm$ 0.07 		&	1730 $\pm$ 195	&  	989 $\pm$  18 \\
H$\gamma$ 	4340.46	&	-1.68	 $\pm$ 0.08		&	1881	 $\pm$  62	&  	950 $\pm$  52 \\
H$\beta$ 	4861.33		& 	-1.55 $\pm$ 0.07		&	1404 $\pm$ 58		&	843 $\pm$ 20 \\		 
H$\alpha$ 6562.80		&	-6.30 $\pm$ 0.10		&	1567 $\pm$ 28		& 	918 $\pm$  53	\\
P$\epsilon$ 9545.972	&		--				& 	--				&		--		\\
Pa$\delta$ 	10049.37	&		--				&	--				&		--		\\
Pa$\gamma$  10938.10	&	-12.40 $\pm$ 0.5		&	1660 $\pm$ 89		& 	--			\\
Pa$\beta$	 12818.08	&	--					&	--				&	--			\\	
Br$\gamma$	21655.29	&	--					&	--				&	--			\\
\hline		
He~{\sc ii} 4025.60		&	-0.666 $\pm$ 	0.12 		& 	1939 $\pm$  155  	&       1062 $\pm$ 67 \\
He~{\sc ii} 4541.59 		&	B				&	--		&	--\\
He~{\sc ii}	 4685.71		&	-3.88  $\pm$  0.22 		& 	1785 $\pm$ 26  	& 	 1022 $\pm$ 42 \\
He~{\sc ii}	 5411.53		&	--					&					& 1167 $\pm$ 17	 \\
He~{\sc ii}	 6683.20		&	-1.88 $\pm$ 0.14		& 	1537 $\pm$ 147	& 	--		\\
He~{\sc ii}	 8236.79		& 	   		--				&		--		&	--		\\
He~{\sc ii}  8801.419  	 	&	B	&--	&--\\
He~{\sc ii}  9225.32		&	B		&	--		&	-- \\
\hline
He~{\sc i} 	4143.76		&	-0.18	 $\pm$ 0.04	&	--					&	--	\\
He~{\sc i}	4471.47 		&	B		&	--		&	--\\
He~{\sc i} 	4921.93		&	-0.57	 $\pm$ 0.05	&	--					&	 --	\\	
He~{\sc i} 	7065.71		&	-0.61 $\pm$ 0.14		&	--				&	--	\\
He~{\sc i} 	8750.47		&	B	&	--	&	--\\
He~{\sc i}	 9210.05		& 	B 		&	--		&	--\\
He~{\sc i} 	10829.09		&	-17.69 $\pm$ 1.30		&	1623 $\pm$ 22		&	943 $\pm$ 8 \\ 
C~{\sc iii}/N~{\sc iii}    4640	&	-1.05 $\pm$ 0.10		&	2639 $\pm$ 158	& -- \\	
\hline
\end{tabular}}
\end{table}

\begin{table}
\centering
\caption{Detected interstellar lines and diffuse interstellar bands
  with their measured equivalent width.}
\medskip
{
\begin{tabular}{llllccc}
\hline
\hline
Line		 			 	&	EW 	 	\\
(\AA)						&	(\AA)	\\			
\hline
Ca~{\sc ii} $\lambda3933.66$		 &	0.41  $\pm$	0.02		\\
Ca~{\sc ii} $\lambda3968.47$		& 	0.30  $\pm$	0.02		 \\
Na~{\sc i}  $\lambda5889.95$		&	0.60 $\pm$	0.06		 \\	
Na~{\sc i}  $\lambda5895.92$		&	0.58 $\pm$	0.01		\\
K~{\sc i}   $\lambda7699.0$		&	0.23  $\pm$	0.01		\\	
\hline
DIB  $\lambda4427$ 		& 		2.69  $\pm$ 0.28\\
DIB  $\lambda5710$ 		&		0.19  $\pm$ 0.02\\
DIB	 $\lambda5781$ 		&		0.51  $\pm$ 0.06\\
DIB  $\lambda5797$ 		&		0.60  $\pm$ 0.06 \\
DIB  $\lambda6283$		& 		1.41  $\pm$	0.02\\
DIB	 $\lambda6614$		&		0.15  $\pm$  0.02\\
\hline
\end{tabular}}
\end{table}

\end{document}